\documentclass[3p,times]{elsarticle}
\pdfoutput=1 
\usepackage{ecrc}
\usepackage{graphicx}  

\newcommand{\pt}           {\ensuremath{p_{\rm T}}}

\newcommand{\rpa}          {\ensuremath{R_{\rm pA}}}
\newcommand{\PbPb}         {\mbox{Pb--Pb}}
\newcommand{\dNdeta}       {\ensuremath{\mathrm{d}N_\mathrm{ch}/\mathrm{d}\eta}}
\newcommand{\snn}          {\ensuremath{\sqrt{s_{\rm NN}}}}
\newcommand{\s}            {\ensuremath{\sqrt{s}}}
\newcommand{\Npart}        {\ensuremath{N_\mathrm{part}}}
\newcommand{\rAA}          {\ensuremath{R_\mathrm{AA}}}

\volume{00}

\firstpage{1}

\journalname{Nuclear Physics A}

\runauth{}

\jid{nupha}

\usepackage{amssymb}

\usepackage[figuresright]{rotating}

\begin{document}

\begin{frontmatter}



\dochead{}

\title{Charged particle production in Pb--Pb collisions at the LHC\\with the ALICE detector}


\author{M. Floris for the ALICE collaboration}
\ead{michele.floris@cern.ch}

\address{CERN, Geneva, Switzerland}

\begin{abstract}
The ALICE collaboration measured charged particle production in $\sqrt{s_{NN}}=2.76$~TeV Pb--Pb collisions at the LHC. We report on results on charged particle multiplicity and transverse momentum spectra.
All the results are presented as  a function of the centrality of the
collision, estimated with a Glauber Monte Carlo fit to multiplicity
distributions reconstructed in various detectors. The applicability of
the Glauber model at LHC energies, the precision of the centrality
determination and the related systematic uncertainties are discussed in detail.

Particles are tracked in the pseudorapidity window $\left|\eta\right| \lesssim 0.9$\ with the silicon Inner Tracking System (ITS) and the Time Projection Chamber (TPC), over the range $0.15 < \pt \lesssim 50$~GeV/$c$. The low-\pt\ cut-off is further reduced in the multiplicity measurement using ``tracklets'', reconstructed in the 2 innermost layers of the ITS.

The charged particle multiplicity is measured in $\left| \eta \right|
< 0.5$ to be $\mathrm{d}N_{ch}/\mathrm{d}\eta = 1601 \pm 60$ in 5\% most central Pb--Pb collisions, indicating an energy density a factor $\sim 3$ higher than at RHIC.
Its evolution with centrality shows a pattern strikingly similar to the one measured at RHIC.
Intermediate ($5 \lesssim \pt \lesssim 15$~GeV/$c$) transverse
momentum particles are found to be most strongly suppressed with respect to pp collisions, consistent with a large energy loss of hard-scattered partons in the hot and dense medium. The results are presented in terms of the nuclear modification factor $R_{\mathrm{AA}}$ and compared to theoretical expectations.
\end{abstract}

\begin{keyword}
heavy-ion \sep charged particles \sep LHC \sep Nuclear modification factor
\end{keyword}

\end{frontmatter}


\section{Introduction}
\label{sec:intro}

The ALICE experiment took data at the LHC in pp collisions at \s~=~0.9,
2.76 and 7~TeV and in \PbPb\ collisions at \snn~=~2.76~TeV.
Measurements of charged particles are very important early studies which
allow for a first characterization of the system produced in such
collisions and provide basic constraints to theoretical models.
In this paper we present results obtained in \PbPb\ collisions for
charged primary particles, defined as all charged particles produced
in the collision, including their decay products, but excluding
products of weak decays of strange particles.

The ALICE detector is mainly composed of a central barrel and of a forward muon
spectrometer~\cite{Alessandro:2006yt}. The main tracking elements in
the central barrel are  a large volume Time Projection Chamber (TPC) and a
six-layered silicon detector, the Inner Tracking System (ITS). They are embedded in a 0.5~T
solenoidal field and cover the pseudorapidity window $\left|\eta\right|
\lesssim 0.9$. The
material budget for a track going through the TPC is about 10\%
of a radiation length. The small material budget and moderate field lead to
a low \pt\ cut-off of about 150 MeV/$c$ for full tracks reconstructed
with the combined information of the ITS and TPC~\cite{Aamodt:2008zz}.
The \pt\ resolution in the present analysis is about 10\% up to
\pt~=~50~GeV/$c$. 
The resolution on the transverse impact parameter relative to the
primary vertex is about 200~$\mu$m at \pt~=~300~MeV/$c$ and 35~$\mu$m
at \pt~=~5~GeV/$c$. This allows for a good separation of primary and
secondary particles, an important requirement for charged particle
measurements at low \pt.
The multiplicity measurements employed  track segments made of 2 hits in the 2
SPD layers pointing to a common vertex (``tracklets'') as the default estimator~\cite{Aamodt:2010cz,Aamodt:2010pb}.
As compared to full-fledged tracks, the tracklets have the advantage of a lower \pt\ cutoff ($\sim
50$~MeV/$c$) and of an extended $\eta$ coverage ($\left|\eta\right| \lesssim 2$).
For triggering purposes, two forward scintillator
hodoscopes, the VZERO detectors, covering the pseudorapidity windows
$2.8 < \eta < 5.1$ and $-3.7 < \eta <
-1.7$ were used. The present analyses use data collected with a minimum
bias trigger requiring a combination of hits in the two innermost
layers of the ITS (made of silicon pixel detectors, SPD) and in the
VZERO.


\section{Centrality determination}
\label{sec:centr}


The determination of the ``centrality'' of a collision, that is a
measurement of the transverse overlap of the two nuclei, is one of the
first important steps towards a full characterization of heavy-ion
collisions. The data are divided in several centrality bins
corresponding to well-defined percentiles of the total hadronic
cross-section. The bins are defined using cuts on the multiplicity
distributions measured in one of the sub-detectors. Our main estimate
is based on the VZERO detector information~\cite{centrality-paper}. Other
estimators are used as cross-checks: clusters reconstructed in the
SPD, tracks reconstructed in the TPC and energy deposited in the Zero
Degree Calorimenters (ZDC, see below).  In order to translate a cut in
any of the above estimators into a given centrality percentile, the
knowledge of the total hadronic cross-section is needed. This can be
extracted from the data in two ways: \textit{i)} fitting the
experimental distribution with a Glauber
model~\cite{Miller:2007ri,Glauber:2006gd}, \textit{ii)} correcting the
experimental distribution for the trigger efficiency and subtracting
any background from non-hadronic interactions. The two approaches
yield consistent results, with the former used as the default.  The
Glauber fit was done using a Monte Carlo calculation, which assumed
Woods-Saxon nuclear density profiles (with parameters constrained by
low energy electron-nucleus scattering measurements) and a
nucleon-nucleon inelastic cross-section of $\sigma^{\rm inel}_{\rm NN}
= 64 \pm 5~\rm mb$, as interpolated from cosmic rays and pp data at
different energies. This value is also consistent with the
measured $\sigma^{\rm inel}_{pp}$~\cite{cross-section:2012he}. From the Glauber model calculation, one can extract
the number of participants $N_{\rm part}$ and the number of binary
collisions $N_{\rm coll}$. It is then assumed that the number of
particle production centers (``ancestors'') is related to the number
of participants and binary collisions as $N_{\mathrm{ancestors}} = f
\cdot N_{\mathrm{part}} + (1-f)\cdot N_{\mathrm{coll}}$. Each ancestor
then produces particles according to a Negative Binomial Distribution
(NBD). The NBD parameters and the $f$ parameter are the free
parametres in the
fit. 

The ZDCs are a set of forward hadronic calorimeters, placed on either
side of the detector, approximately 114~m away from the interaction
point, at zero degree with respect to the beam axis. For the most
central events, the energy deposit in the ZDCs is proportional to the
number of spectator nucleons. They therefore measure a quantity
qualitatively different from the other estimators, providing another
implicit confirmation of the applicability of the Glauber model at the
LHC.

Comparing the different estimators event-by-event, it is possible to
estimate the precision of the centrality measurement. For the VZERO
estimator, the centrality resolution is $\sim0.5\%$ for central events
and $\sim2\%$ for peripheral events.  The residual contamination and
purity of the events used in the analysis was also studied. This is
particularly important for peripheral events.  A pure hadronic sample
is found after the basic offline selection down to about 90\%
centrality~\cite{centrality-paper}.  The minimum bias trigger
efficiency ranges between 97 and 99\%, according to the actual trigger
settings. More details on the centrality studies can be found
in~\cite{centrality-paper}.

\section{Multiplicity measurements}
\label{sec:mult-meas}

\begin{figure}[tb]
\centering  
\includegraphics[width=0.40\textwidth]{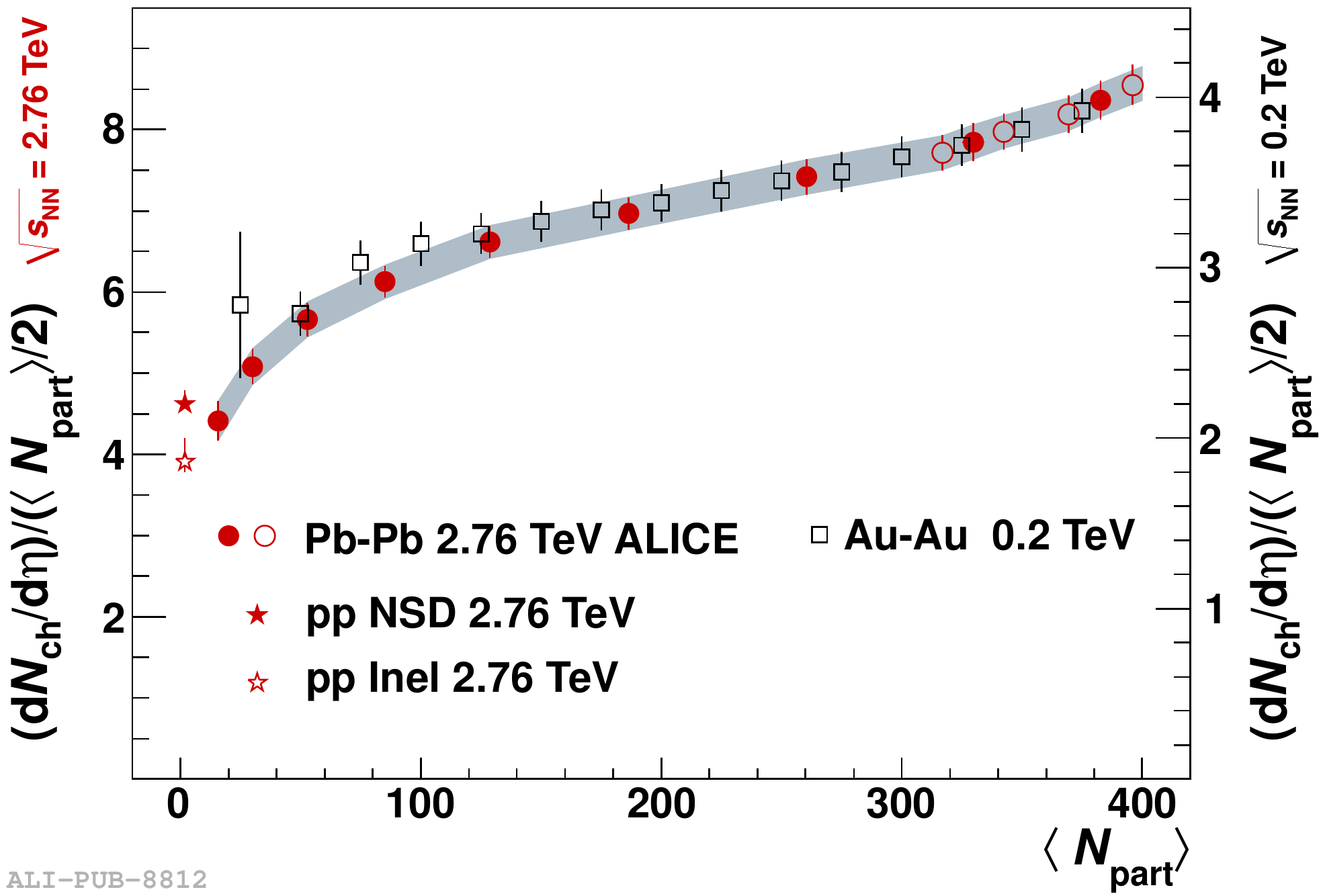}  
\hfill
\includegraphics[width=0.38\textwidth]{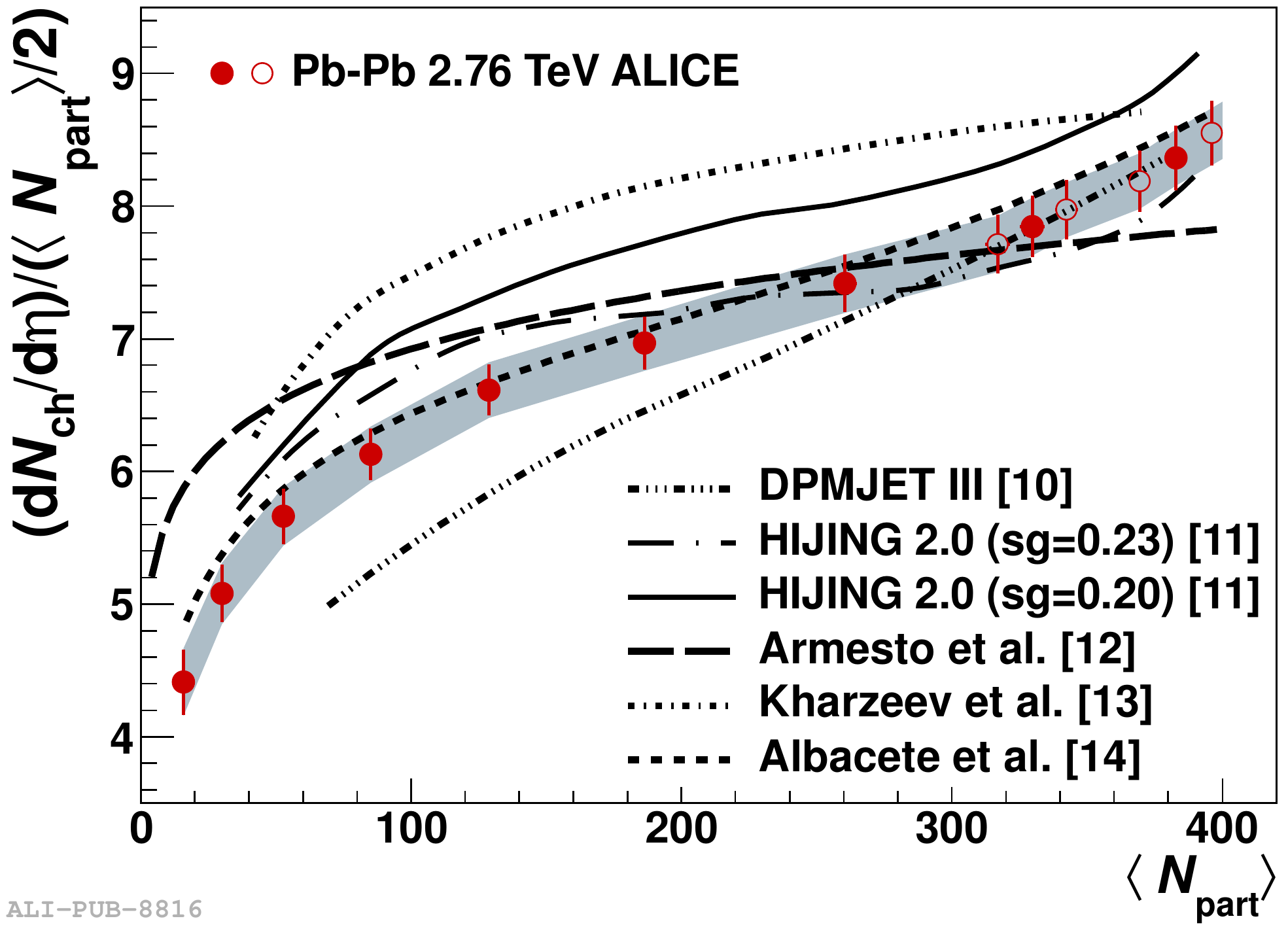}

\caption{Charged particle pseudorapidity density per
  participant pair \dNdeta/(\Npart/2) measured by ALICE, as a function of the number of
  participants \Npart, compared to previous RHIC results (scaled by a
  factor 2.14, left) and theoretical models
  (right).}
\label{fig:mult}
\end{figure}

The charged particle multiplicity was measured in the central rapidity
region ($\left| \eta \right| < 0.5$) using ``tracklets'' as the
default estimator and crosschecked with tracks. The \dNdeta\ as a
function of \Npart\ is compared in Fig.~\ref{fig:mult} with previous
RHIC results at \snn~=~200~GeV/$c$ (left) and with theoretical models
(right, see~\cite{Aamodt:2010cz} and references therein). The
dependence on centrality is very similar at the two energies, with an
overall scaling factor of $\sim 2.14$.  Models incorporating a
moderation of the multiplicity with centrality (i.e. shadowing or
saturation) are favored by the data.  The measurement has recently
been extended to cover a much wider pseudorapidity range ($\left|\eta\right|
\lesssim 5$)~\cite{maxime-QM}, putting further constraints on
theoretical models. The future measurement in p--A collisions, foreseen
at the beginning of 2013 with the first p--A run at the LHC, will also
be of great importance for constraining the initial conditions for
nuclear collisions at LHC energies.

\section{High \pt\ particle suppression}
\label{sec:high-pt-particle}

The study of single inclusive charged particle spectra allows to
investigate the possible suppression of high \pt\ hadrons, due to
parton energy loss in the medium.
The production (and suppression) of high \pt\ particles was studied in
terms of the nuclear modification factor \rAA, defined as the ratio of
the yield in \PbPb\ collisions to the pp yield scaled by the number
of binary collisions. This ratio can also be written as:
\[
\rAA = \frac{d^2N/\mathrm{d}\pt\mathrm{d}\eta}{\langle
  T_{\mathrm{AA}}
  \rangle\mathrm{d^2}\sigma_{pp}^{\rm inel}/\mathrm{d}\pt\mathrm{d}\eta} ,
\]

where $T_{\mathrm{AA}}$ is the nuclear overlap function computed from the
Glauber model and $\sigma^{\rm inel}_{pp}$ the pp inelastic cross section.
The pp reference is a crucial ingredient for the calculation of the
\rAA. It is based in the present analysis on a direct measurement in
pp collisions at
\s~=~2.76~TeV. This allows for a reduction of the systematic
uncertainty and increased \pt\ range with respect to the first
published results~\cite{Aamodt:2010jd}. The pp measurement extends up to
$\pt~=~35$~GeV/$c$. In order to reduce the effect of the statistical
fluctuations and to extend the \pt\ coverage, it was
fitted to a
modified Hagedorn function and extrapolated to higher \pt~\cite{raa-paper,Otwinowski:2011gq}.

\begin{figure}[tb]
\centering  
\includegraphics[width=0.40\textwidth]{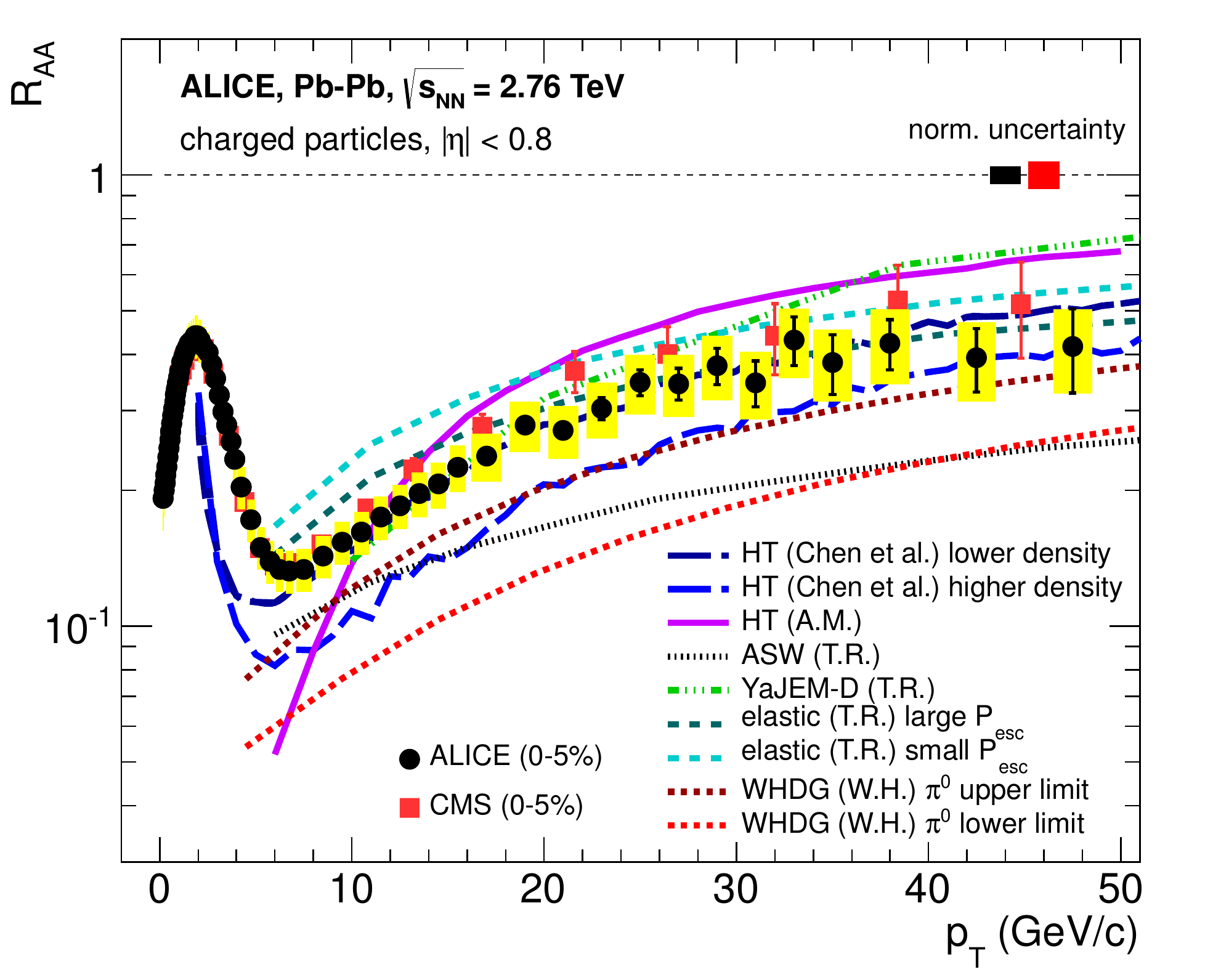}
\hfill
\includegraphics[width=0.38\textwidth,trim=0 366 0 0,clip=true]{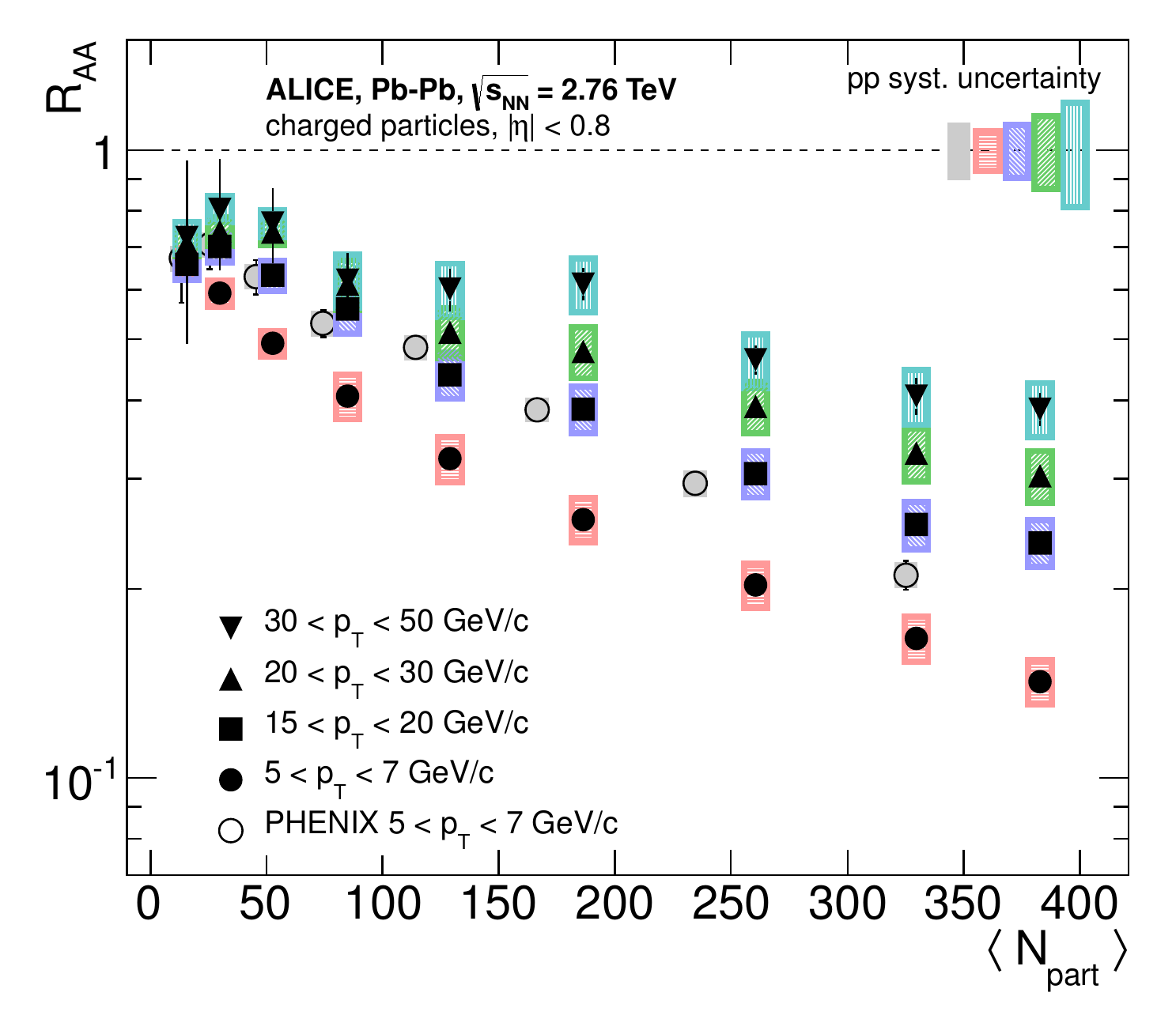}  
  \caption{Nuclear modification factor \rAA\ for the 5\% most central
    collisions, compared to theoretical calculations and to CMS data
    (left); Centrality dependence of \rAA\ in four \pt\ windows,
    compared to PHENIX results in Au--Au collisions at \snn~=~200~GeV (right).}
\label{fig:raa}
\end{figure}

The \rAA\ as a function of \pt\ for central collisions is shown in Fig.~\ref{fig:raa}
(left)\footnote{In this paper the final results, which became
  available after the conference presentation~\cite{raa-paper}, are presented.}. A strong
suppression is observed, with the \rAA\ showing a pronounced minimum
at $\pt \simeq 6$~GeV/$c$.
At lower \pt, the \rAA\ rises and develops a peak at $\pt \simeq
2$~GeV/c. This can be understood in terms of hydrodynamic
flow~\cite{Muller:2012zq}.
In the range $2 \lesssim \pt \lesssim 6$~GeV/$c$ the \rAA\ decreases,
reaching the mimimum mentioned above. This intermediate \pt\ region is
likely driven by an interplay of soft and hard processes. 
Above $\pt\simeq 6$~GeV the \rAA\ shows a rise, which becomes
progressively less steep, possibly reaching a saturation. The \rAA\ is
also compared in Fig.~\ref{fig:raa} (left) to a similar measurement by
the CMS experiment~\cite{CMS:2012aa}. The results from the two
experiments are consistent with each other~\cite{raa-paper}.
The right panel of Fig.~\ref{fig:raa} shows the centrality dependence
of the \rAA\ in different \pt\ regions, as a function of \Npart. The
data are also compared to previous measurements by the PHENIX
experiment in Au--Au collisions at \snn~=~200~GeV~\cite{Adler:2003au} for the region around
the minimum ($4 < \pt <7$~GeV/$c$). The suppression increases with
\Npart\ (and hence with centrality) at all \pt. The strongest
centrality dependence is observed for the region around the
minimum~\cite{raa-paper}. As compared to the lower energy result we observe that the
suppression is stronger at the LHC at all centralities (expressed in
terms of \Npart). If central RHIC results are compared to LHC results
at a similar $\dNdeta \simeq 700$, however, it is found that the
suppression is comparable at the 2 energies.

The comparison with the theoretical calculations depicted in
Fig.~\ref{fig:raa} (left)~\cite{Horowitz:2011gd,Salgado:2003gb,Chen:2011vt,Majumder:2010ik,Renk:2011gj}
shows that many models can reproduce the trend and level of suppression
above $\pt \simeq 6$~GeV/$c$. Due to the uncertainties in several of
the input parameters to the model, different approaches are
reproducing in a satisfactory way the \rAA\ results.
The uncertainty on the initial conditions will
be better constrained with the upcoming p--A run at the LHC, with a
measurement of the nuclear modification factor \rpa\ and a measurement
of the charged particle multiplicity over a large pseudorapidity range.
Additional constraints~\cite{Horowitz:2011gd} on the models come from the
recently-released ALICE results on
identified light flavour hadrons up to \pt~=~20~GeV/$c$~\cite{Antonio-QM}.







\bibliographystyle{elsarticle-num}
\bibliography{hp2012}







\end{document}